%% file: main.tex
\documentclass[conference]{IEEEtran}
\IEEEoverridecommandlockouts
\usepackage{cite}
\usepackage{amsmath,amssymb,amsfonts}

\usepackage{algorithm}
\usepackage[noend]{algorithmic}
\usepackage{xcolor}
\usepackage{tikz}
\usetikzlibrary{fit,calc}

\colorlet{pink}{red!40}

\usepackage{ctable}

\usepackage{graphicx}
\usepackage{textcomp}
\usepackage{xcolor}
\usepackage{soul}
\usepackage{multirow}
\usepackage[normalem]{ulem}
\def\BibTeX{{\rm B\kern-.05em{\sc i\kern-.025em b}\kern-.08em
    T\kern-.1667em\lower.7ex\hbox{E}\kern-.125emX}}
\begin{document}

\title{G-CoS: GNN-Accelerator Co-Search Towards Both Better Accuracy and Efficiency
}

\author{
\IEEEauthorblockN{Yongan Zhang\textsuperscript{1},
Haoran You\textsuperscript{1}, Yonggan Fu\textsuperscript{1}, 
Tong Geng\textsuperscript{2}, Ang Li\textsuperscript{2}, Yingyan Lin\textsuperscript{1}}
\IEEEauthorblockA{
\textsuperscript{1}Rice University, Houston, TX; 
\textsuperscript{2}Pacific Northwest National Laboratory, Richland, WA\\
Corresponding to:
\textsuperscript{1}yingyan.lin@rice.edu}\vspace*{-1cm}}
\maketitle

\input{Sections/0-Abstract}
\input{Sections/1-Introduction}

\input{Sections/2-Related_work}

\input{Sections/3-Method}

\input{Sections/4-Experiment}
\input{Sections/5-Conclusion}

\bibliography{ref}
\bibliographystyle{IEEEtran}

\end{document}

%% file: Sections/0-Abstract.tex
\begin{abstract}
Graph Neural Networks (GNNs) have emerged as the state-of-the-art (SOTA) method for graph-based learning tasks. However, it still remains prohibitively challenging to inference GNNs  over large  graph  datasets, limiting  their  application to large-scale real-world tasks. While end-to-end jointly optimizing GNNs and their accelerators is promising in boosting GNNs’ inference efficiency and expediting the design process, it is still underexplored due to the vast and distinct design spaces of GNNs and their accelerators. In this work, we propose G-CoS, a GNN and accelerator co-search framework that can  automatically search for matched GNN structures and accelerators to maximize both task accuracy and acceleration efficiency. Specifically, G-CoS integrates two major enabling  components: (1) \textit{a generic GNN accelerator search space} which is applicable to various GNN structures and (2) \textit{a one-shot GNN and accelerator co-search algorithm} that enables simultaneous and efficient search for optimal GNN structures and their matched accelerators. To the best of our knowledge, G-CoS is the first co-search framework for GNNs and their accelerators. Extensive experiments and ablation studies show that the GNNs and accelerators generated by G-CoS consistently outperform SOTA GNNs and GNN accelerators in terms of both task accuracy and hardware efficiency, while only requiring a few hours for the end-to-end generation of the best matched  GNNs and their accelerators. 

\end{abstract}


%% file: Sections/1-Introduction.tex
\section{Introduction}
\label{sec:intro}

Graph neural networks (GNNs) \cite{kipf2016semi} have gained an increased popularity recently as they demonstrated the state-of-the-art (SOTA) performance in various graph-based learning tasks, including node classification \cite{kipf2016semi}, 
graph classification \cite{xu2018powerful}, and recommendation systems \cite{ying2018graph}. However, GNNs often suffer from prohibitive inference cost, limiting their potential to handle large-scale real-world graph applications.
For example, a 2-layer GNN model requires 19G FLOPs (FLOPs: floating point operations) to inference the Reddit graph \cite{tailor2021degreequant}, which requires a latency of $2.94$E+$5$ milliseconds when being executed on an Intel Xeon E5-2680 CPU platform \cite{geng2020awb}, i.e., its required FLOPs and latency are 2$\times$ and 5000$\times$ of a 50-layer convolutional neural network (CNN), ResNet-50 \cite{res50_latency}.

The giant computational cost of GNN inference results from three aspects. 
\underline{First}, graphs are often very large as exacerbated by their intertwined complex neighbor connections, e.g., a total of 232,965 nodes in the Reddit graph with each node having about 50 neighbors \cite{hamilton2017inductive}. 
\underline{Second}, real-world graphs tend to follow the power-law distribution and therefore have highly irregular adjacent matrices, resulting in prohibitive inefficiencies in both data processing and movements.
\underline{Third}, the dimension of GNNs' node feature vectors can be very high, e.g., each node in the CiteSeer graph has 3703 features.

To tackle GNNs' prohibitive inference cost, various efficient GNN inference techniques have been developed. On the algorithm level, several pioneering GNN compression techniques have been developed. For instance, two concurrent GNN pruning works \cite{li2020sgcn,zheng2020robust} aim to sparsify the connections in GNNs' graph adjacent matrices; and \cite{tailor2021degreequant} for the first time shows the feasibility of adopting 8-bit integer arithmetic for GNN inference without sacrificing the accuracy. Another paralleled trend is to search for efficient GNN architectures \cite{gao2019graphnas,design_space}. On the hardware level, various GNN accelerators have been proposed. For example, HyGCN \cite{yan2020hygcn} proposes hybrid execution patterns of GNNs for leveraging their intra-vertex and inter-vertex parallelisms to handle the irregularity in the aggregation phase and reusability in the combination phase, respectively;
Later, AWB-GCN \cite{geng2020awb} identifies the workload imbalance problem in the aggregation phase, and proposes auto-tuning workload balancing techniques, achieving an average speedup of 7.4$\times$ over HyGCN. On the development tool level, pioneering works have attempted to characterize the design space of dataflows and micro-architectures for GNN accelerators \cite{garg2021taxonomy}, and develop an automated framework to generate GNN accelerators \cite{liang2020deepburning}. 

Despite the promising performance of existing efficient GNN inference solutions, their achievable efficiency is still not sufficient for enabling extensive GNN inference applications due to GNNs' extremely dynamic and irregular data accesses and thus excessive acceleration cost. Motivated by the great success of algorithm-accelerator co-exploration works for CNN accelerations~\cite{DIAN,edge_fpga_co_design, abdelfattah2020best, fu2021autonba, Zhang2021RTRCGNN, jiang2019hardware, li2020edd}, this work targets \textcolor{black}{to co-optimize both the GNN structures and their accelerators with boosted development speed}, and makes the following contributions:

\begin{itemize}

\item We propose G-CoS, a GNN and accelerator co-search framework that can  automatically search for the matched GNN structures and accelerators to maximize both task accuracy and acceleration efficiency. To the best of our knowledge, G-CoS is the first co-search framework for GNNs and their accelerators. 

\item G-CoS integrates two enabling  components: (1) \textit{a generic GNN accelerator search space} which is applicable to various GNN structures and (2) \textit{a one-shot GNN and accelerator co-search algorithm} that enables simultaneous and efficient search for optimal GNN structures and their matched accelerators, both of which can facilitate the algorithmic exploration of efficient GNN solutions. 

\item Extensive hardware/algorithm experiments and ablation studies validate G-CoS's effectiveness and advantage: G-CoS generated networks/accelerators consistently outperform SOTA GNNs/accelerators, while requiring only a few hours for the end-to-end search, \textcolor{black}{(e.g., 4 GPU hours for the Cora dataset)}. 


\end{itemize}

%% file: Sections/2-Related_work.tex
\section{Related Works}

\textbf{Graph neural networks (GNNs).}
GNNs have achieved great success in graph-based learning tasks \cite{zhang2018end,gori2005new}. Depending on their graph representation domains, GNNs can be categorized into spectral and spatial GNNs. Specifically, spectral GNNs model the representation in the graph Fourier transform domain based on eigen-decomposition and usually handle the whole graph simultaneously \cite{kipf2017semi, peng2020learning}. However, it becomes impractical for spectral GNNs to process large graphs and difficult for them to take advantage of parallel processing \cite{gao2019graphnas, wu2020comprehensive}. On the other hand, spatial GNNs \cite{hamilton2017inductive,simonovsky2017dynamic}, which directly perform the computation in the graph domain by aggregating the neighbor nodes’ features, have undergone rapid development. Recently, \cite{GAT} introduced an attention mechanism to further improve the performance of spatial GNNs; and \cite{zeng2019accurate} utilized mini-batch training to improve GNNs' scalability to handling large graphs. Combined with sampling strategies, the whole graph is no longer required during aggregation, leaving much room for potential hardware acceleration \cite{hamilton2017inductive, gao2018large}. Therefore, in this work, we will primarily focus on the spatial GNNs for their advantages on scalability and potential hardware acceleration.

\textbf{Graph neural architecture search (GNAS).}
Neural architecture search (NAS) has become a popular approach to designing neural networks~\cite{zoph2016neural, pham2018efficient,bello2017neural}, which can significantly relieve human efforts from manually designing complex networks. The recent NAS success and the large distinction among GNN structures for different tasks have motivated the use of NAS for GNNs (denoted as GNAS). For example, \cite{gao2019graphnas, zhou2019auto} used reinforcement learning (RL) methods along with parameter sharing to efficiently search for GNNs; \cite{kyriakides2021evolving} adopted an evolutionary search algorithm; and \cite{you2020} proposed a more generic GNN design space and a standardized evaluation method for GNNs across various graph learning tasks. Despite the preliminary success, existing works still heavily rely on excessive rounds of sampling and retraining, limiting their scalability to more generic search spaces.


\textbf{Hardware-aware architecture search (HA-NAS).}
To ensure the searched networks' hardware efficiency, hardware-aware NAS (HA-NAS) proposes to incorporate hardware metrics, e.g., the latency on mobile phones, into the search process. Early works, e.g., \cite{tan2019mnasnet, howard2019searching, tan2019efficientnet}, utilized RL-based methods, and thus suffered from substantial search time and costs, limiting their scalability to larger and more diverse search spaces. Inspired by DARTS~\cite{liu2018darts}, differentiable HA-NAS~\cite{wu2019fbnet, wan2020fbnetv2, jin2019rc, li2020edd} has emerged to greatly improve both the search and hardware efficiency. 
However, restricted by the large difference among different GNN structures and thus the difficulty for hardware acceleration, HA-NAS targeting GNNs has rarely been explored. Furthermore, existing HA-NAS methods have not yet fully explored the hardware design space. As the acceleration efficiency is determined by both the network structures and their accelerators, it is thus desirable to jointly search for both the networks and their accelerators.  

\textbf{GNN inference accelerators.}
GNNs' ultra-sparse graph matrices, corresponding to extremely dynamic and irregular data accesses as well as distinct execution patterns from DNNs, have fueled a growing interest in developing dedicated GNN accelerators \cite{autenhardware}. For instance, HyGCN \cite{yan2020hygcn} explored both intra/inter-vertex parallelisms to separately handle the irregularity in the aggregation phase and reusability in the combination phase.
Later, aiming to boost the overall hardware utilization, AWB-GCN \cite{geng2020awb} proposed to balance the workload during runtime with an auto-tuning algorithm and to increase the data locality by regionally clustering the non-zero values (i.e., connected neighbors) within the adjacency matrices; EnGN \cite{liang2020engn} proposed a ring-edge-reduce dataflow to handle graphs with arbitrary dimensions and increase the accelerator's scalability to large graphs; and GRIP \cite{kiningham2020grip} employed fine-grained vertex-tiling to reduce the weight bandwidth requirements;
In parallel, to reduce the human efforts in designing GNN accelerators and democratize the process, pioneering works have attempted to characterize the design space of dataflows and micro-architectures for GNN accelerators \cite{garg2021taxonomy}, and developed an automated framework to generate GNN accelerators \cite{liang2020deepburning}. Nevertheless, existing automated frameworks for GNNs still have limited support to various GNN structures and thus suffer from low hardware utilization and achievable efficiency on certain tasks.


\textbf{Software/Hardware Co-exploration.} 
Jointly exploring the networks and their accelerators has shown promising results~\cite{DIAN,edge_fpga_co_design, abdelfattah2020best, fu2021autonba, Zhang2021RTRCGNN, jiang2019hardware, li2020edd,zhao2020smartexchange}. 
For instance,~\cite{edge_fpga_co_design, jiang2019hardware} conducted RL-based search to jointly optimize the networks and some design parameters of FPGA-based accelerators; ~\cite{DIAN} developed the first differentiable network and accelerator co-search framework to boost both the task accuracy and acceleration efficiency; and~\cite{fu2021autonba} co-searches for networks, bitwidths, and accelerators to achieve superior performance. However, co-optimizing the GNN structures and their accelerators has not been studied.

\section{GNN preliminaries}

\subsection{GNN notation and formulation} 
A typical GNN graph can be represented as, $G = (V, E)$, where $v_i \in V$ and $(v_i, v_j) \in E$  denote the nodes and edges, respectively; and $N = | V |$ and $M = | E |$ denote the total number of nodes and edges, respectively. The node degree is denoted as $d = \{d_1, d_2, \cdots, d_N\}$ where $d_i$ indicates the number of neighbors connected to node $v_i$. We define $D$ as the degree matrix whose diagonal elements are formed using $d$. The connectivity information is encoded within the adjacency matrix $A \in \mathbb{R}^{N \times N}$, where the non-zero entries represent the existed connections among different nodes. For each layer $l$ of a GNN, the nodes are encoded by their feature vectors $\{x_1^{(l)}, x_2^{(l)}, \cdots, x_N^{(l)}\} = X^{(l)}$, where $X^{(l)} \in \mathbb{R}^{N\times F_{(l)}}$ and $F_{(l)}$ denotes the feature dimension used to encode the nodes at layer $l$.  Thus, a GNN layer \cite{kipf2017semi} can be formulated as:

\begin{equation}\label{eq:gcn}
    X^{(l+1)} = \text{ACT}_{(l)} \left(\hat{A}X^{(l)}W^{(l)}\right) ,
\end{equation}
where $\hat{A}$ is a normalized version of $A$: $\hat{A}=D^{-\frac{1}{2}} A D^{-\frac{1}{2}}$, $\text{ACT}_{(l)}$ represents the activation function of layer $l$ and $W^{(l)}\in \mathbb{R}^{F_{(l)}*K_{(l)}}$ represents the weights in layer $l$, with $K_{(l)}$ denoting the hidden/weight dimension at layer $l$. 
The whole GNN inference can thus be viewed as two separated phases: \textit{Aggregation} and \textit{Combination}.

\begin{itemize}
\itemsep -0.3\parsep
    \item \textit{Aggregation} $\hat{A}X^{(l)}$: For each node in the graph, a GNN aggregates its 1-hop neighbor nodes' features into a unified feature vector, corresponding to the multiplication between the adjacent and feature matrix, i.e., $\hat{A} X$.
    \item \textit{Combination} $[\hat{A}X^{(l)}] W^{(l)}$: The aggregated feature vector, $\hat{A}X^{(l)}$, will be further transformed to another feature vector via an MLP network (shared among nodes) with weights $W^{(l)}$ for learning better representations, corresponding to the multiplication between the aggregated feature matrix and weight matrix, i.e.,  $[\hat{A}X^{(l)}] W^{(l)}$.
\end{itemize}
In the final/prediction layer of GNNs, after the feature vectors' update, a softmax function is usually applied in a row-wise manner, i.e., $\textit{softmax}(x_i^{(l)}) = \text{exp} (x_i^{(l)}) / \sum_i \text{exp} (x_i^{(l)})$  \cite{kipf2017semi}. For semi-supervised multiclass classification, the loss function captures the cross-entropy errors over all labeled examples:
\begin{equation}\label{eq:gcn_loss}
    \mathcal{L}_{GNN}(W) = - \sum_{n \in \mathcal{Y}_N} \sum_f Y_{nf} \, ln(\Theta_{nf}),
\end{equation}
where $\mathcal{Y}_N$ is the set of node indices that have labels, $Y_{nf}$ is the ground truth label matrix, and $\Theta_{nf}$ denotes the predicted possibilities of node $n$ belonging to class $f$. During GNN training, $W^{(l)}$ is updated via gradient descents.

\subsection{GNN variants and their implementations}
Many advanced GNN variants have recently been proposed to consider different aggregation functions and introduce additional attention modules or sampling functions.
Without loss of generality, we summarize four popular GNN architectures:
\textbf{GCN} \cite{kipf2017semi}, \textbf{GAT} \cite{GAT}, \textbf{GIN} \cite{xu2018how}, and \textbf{GraphSAGE} \cite{hamilton2017inductive}. We analyze the difference among them as compared with vanilla GNNs below, \textbf{aiming to generally support them in G-CoS}.

\textbf{GCN~\cite{kipf2017semi}}: During inference, each node can be written as $x_i^{(l+1)} = \sum_{j \in \mathcal{N}(i) \cup i} (\frac{1}{d_i d_j} W^{(l)} x_j^{(l)})$, where $l$ is the layer index and $\mathcal{N}(i)$ represents the $i$-th node's neighbor set. Thus, compared with the vanilla GNNs, the only difference lies in the entries of the adjacency matrix where each node is encoded as $\frac{1}{d_i d_j}$ and \textit{can be filled offline before the processing}.

\textbf{GAT~\cite{GAT}}: An attention module is introduced, i.e., $x_i^{(l+1)} = \alpha_{i,i}W^{(l)}x_i^{(l)} $ $ + \sum_{j \in \mathcal{N}(i)}(\alpha_{i,j}W^{(l)}x_j^{(l)})$, where $\alpha$ denotes the attention coefficients for the neighbor nodes and can be viewed as the elements to replace the original adjacency matrix's entries. Adapting from the formulation of GAT~\cite{GAT, shi2021versagnn}, $\alpha_{i,j}$ can be calculated as: 
\begin{equation}\label{eq:gat}
    \alpha_{i,j}=\frac{exp(\text{ACT}({ x_i^\intercal w_1^{(l)}}+x_j^\intercal{w_2^{(l)}} ))}{ \sum_{k \in \mathcal{N}(i) \cup i}exp(\text{ACT}( x_i^\intercal w_1^{(l)}+x_k^\intercal w_2^{(l)} ))}
\end{equation}
where \text{ACT } denotes the activation used in the attention module and $(w_1^{(l)},w_2^{(l)})\in (\mathbb{R}^{F_{(l)}\times 1}, \mathbb{R}^{F_{(l)}\times 1})$ denotes the weights of the attention module. The whole set of $\alpha$ can then be calculated by introducing  an additional layer of matrix multiplication of $X^{(l)} * [w_1^{(l)} || w_2^{(l)}]$ along with the element-wise activation and multiplication with the original adjacency matrix. Thus, \textit{replacing the original adjacency matrix with $\alpha$} captures the functionality of the attention module.

\textbf{GIN~\cite{xu2018how}}: An information-lossless aggregation function is adopted, i.e., $x_i^{(l+1)} = \text{MLP}((1 + \epsilon)x_i^{(l)} + \sum_{j \in \mathcal{N}(i)} x_j^{(l)})$, where MLP denotes an MLP network and $\epsilon$ is a learnable constant. As  $\epsilon$ is trained and then fixed during inference, GIN can be realized by \textit{fusing $\epsilon$ into the original adjacency matrix and incorporating an additional MLP layer into the aggregation phase of vanilla GNNs}.
    
\textbf{GraphSAGE~\cite{hamilton2017inductive}}: Uniform neighbor sampling is introduced to alleviate the extreme memory consumption during training, i.e., $x_i^{(l+1)} = \textit{Mean} (W^{(l)} x_j^{(l)})$, $j\in \{i\} \cup \{ \mathcal{S}(i)\}$, where $\mathcal{S}(i)$ denotes the sampled neighbors, which can be easily supported by \textit{introducing an additional node sampling layer and using mean aggregation} to the original GNN formulation as in Eq.~\ref{eq:gcn}.

%% file: Sections/3-Method.tex
\begin{figure}[t]
    \centering
    \includegraphics[width=\linewidth]{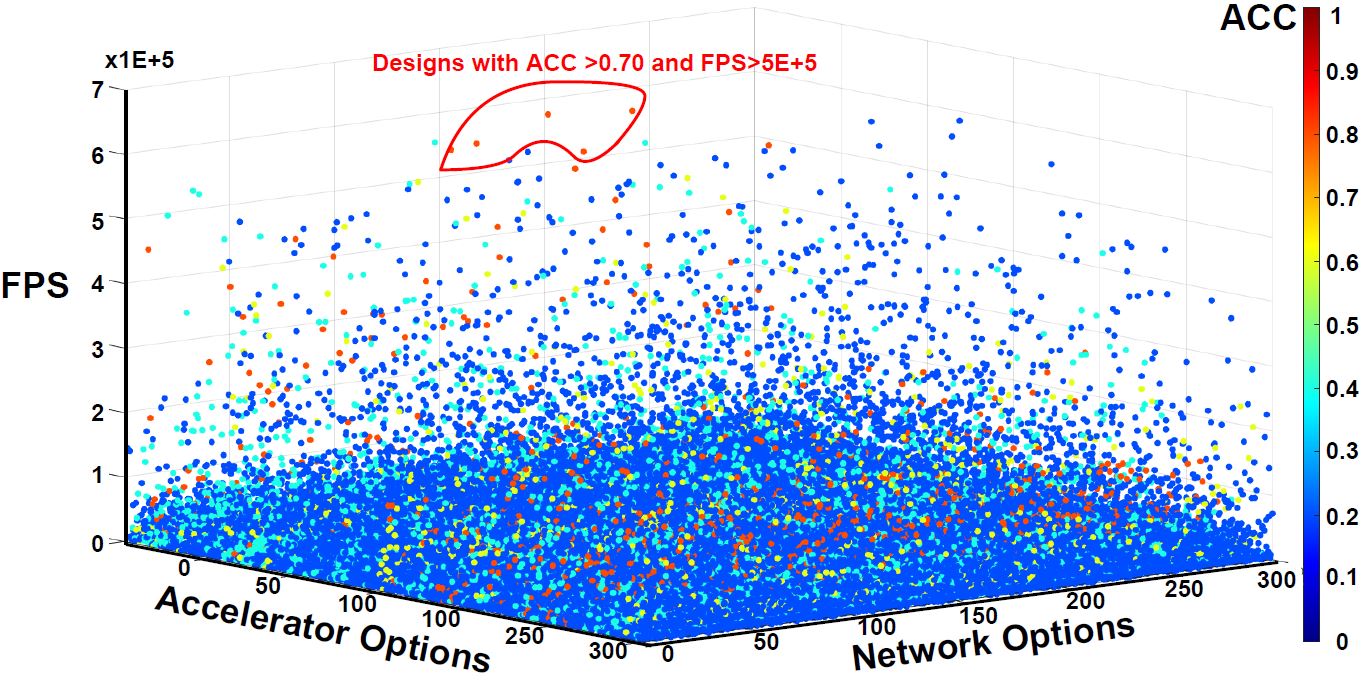}
    \caption{FPGA measured Frame-Per-Second (FPS; see the left axis) on a VCU128 FPGA~\cite{vcu128} and accuracy on Cora dataset (see the right colorbar) of 300 randomly sampled fully trained GNNs from supernet as defined in Sec.~\ref{sec:gnas_space}, when each of the networks is accelerated by 300 randomly sampled accelerators from the accelerator design space (see Sec.~\ref{sec:hw_space}), leading to a total of $9$E+$4$ randomly sampled data points. Designs with $ACC>0.7$ and $FPS >5$E+$5$ are circled out in red, which are extremely sparse.}
    \label{fig:huge_space}
    \vspace{-1.5em}
\end{figure}

\begin{figure*}[t]
    \vspace{-1.5em}
    \centering
    \includegraphics[width=0.85\linewidth]{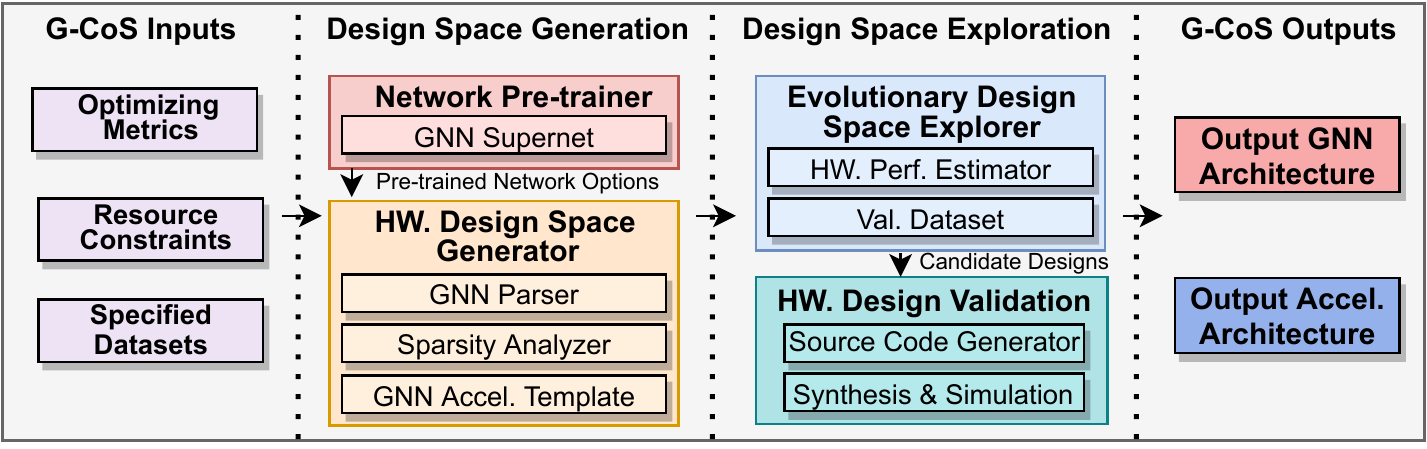}
    \vspace{-1em}
    \caption{An overview of our G-CoS GNN-accelerator co-search framework, where Accel. denotes accelerators. }
    \vspace{-1.5em}
    \label{fig:overview}
\end{figure*}


\section{The Proposed G-CoS framework}
This section describes our G-CoS framework by first providing an overview and the problem formulation, and then G-CoS's two major enablers: a generic GNN accelerator design space and network structure design space, followed by G-CoS's efficient one-shot evolutionary co-search algorithm.

\subsection{G-CoS: the overview and challenges}
\label{sec:overview}

To maximize both task accuracy and hardware efficiency, G-CoS jointly searches for the best matched GNN structures and accelerators, under the specified datasets, resource constraints, and optimizing metrics (e.g., accuracy and latency), as shown in Fig.~\ref{fig:overview}. 

To enable effective GNN-accelerator co-search, there exist \textbf{three major challenges}, including (1) the prohibitively large and distinct joint space versus very sparse optima excelling at both accuracy and efficiency, as shown in Fig.~\ref{fig:huge_space}, (2) the excessive retraining cost during GNAS, and (3) the lack of either generic GNN structure or accelerator search space description. To tackle the first two aforementioned challenges, G-CoS employs \textit{a one-shot evolutionary GNN-accelerator co-search algorithm}, as introduced in Sec.~\ref{sec:G-CoS-alg}. G-CoS first one-shot pre-trains the proposed GNN supernet to avoid the necessity of cumbersome retraining in the GNN-accelerator co-search phase, and then utilizes an evolutionary search algorithm to efficiently navigate the joint network-accelerator space to locate the optimal GNN-accelerator pairs based on the feedback of the estimated inference accuracy and hardware efficiency. For the third challenge, G-CoS integrates (1) \textit{a generic GNN network space} description which is compatible with its one-shot search algorithm and (2) \textit{a generic GNN accelerator design space} which includes accelerators with high hardware utilization across various GNN structures.  

\subsection{G-CoS: the co-optimization formulation}
G-CoS's co-optimization process can be formulated as:
\begin{equation} \label{eq:objective}
    arg\min_{\{gnet, hw\}} L_{val}(\omega^*, gnet, hw) + \lambda L_{cost}(gnet, hw)
\end{equation}
\begin{equation}
    s.t.\quad \omega^*= arg\min_\omega L_{train}(\omega, gnet), 
\end{equation}
where $\omega$ denotes the GNN weights; $L_{train}$, $L_{val}$, and $L_{cost}$ are the task loss during training, task loss during validation, and hardware-cost, respectively, given the GNN structure, the accelerator parameter set, and the specified metrics; and $gnet$ and $hw$ are the selected GNN structure and accelerator to be optimized, respectively. Note that the hardware-cost is co-determined by both the GNN structure and accelerator.

\subsection{G-CoS: a generic GNN accelerator template and space}
\label{sec:hw_space}

To comprehensively cover potential parallelism and reuse opportunities for various GNN structures, we propose a generic GNN accelerator template along with a set of corresponding searchable parameters, forming a design space 
featuring a total of $\sim$ $1$E+$15$ GNN accelerator choices with different micro-architectures and dataflows.

\begin{figure}[t]
    \centering
    \includegraphics[width=\linewidth]{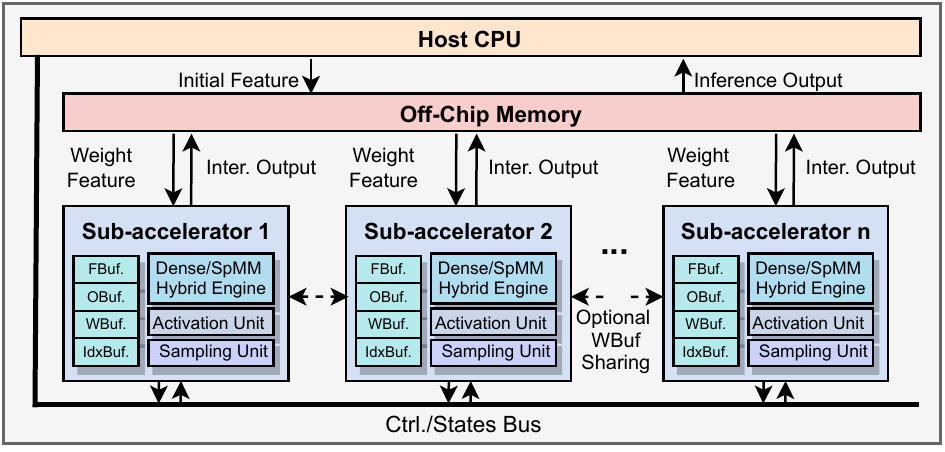}
    \vspace{-2em}
    \caption{An illustration of G-CoS's accelerator micro-architecture template. }
    \label{fig:hw_template_overview}
    \vspace{-2em}
\end{figure}

\begin{figure*}[t]
    \centering
    \vspace{-3em}
    \includegraphics[width=\linewidth]{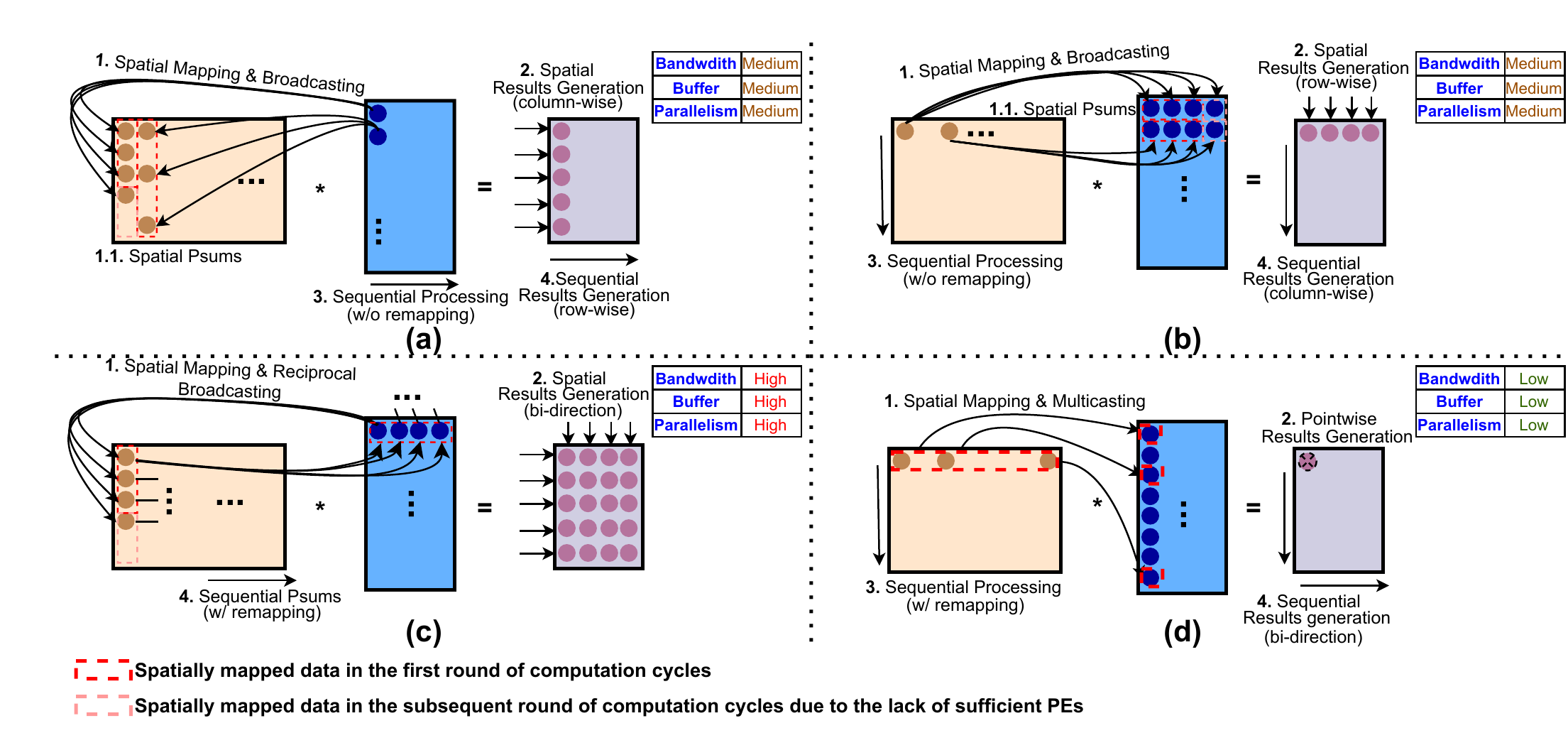}
    \vspace{-3.3em}
\caption{The choices of kernel modes for each sub-accelerator, where different modes represent different spatial mapping/temporal mapping methods. For better visual clarity, the operation order for each mode is numbered in each sub-figure, and the corresponding properties on off-chip bandwidth consumption, on-chip buffer consumption, and potential parallelism opportunities are summarized on the right corner of each sub-figure. }
    \label{fig:comp_mode}
     \vspace{-2em}
\end{figure*}

\textbf{The micro-architecture overview.} 
To maximize the acceleration throughput while at the same time minimizing the latency, we adopt a multi-accelerator micro-architecture template to accelerate both the combination and aggregation phases, as shown in Fig.~\ref{fig:hw_template_overview}. This template has two overall advantages:

\begin{itemize}
    \item \textbf{Latency friendly:} When working on either of the two phases, all the hardware components will be instantiated and subsequently reused for the other phase, thus reducing the startup latency of GNN inference which would otherwise be much higher if a pipeline structure was employed for the two phases as in~\cite{geng2020awb}.
    \item \textbf{Utilization friendly:} Given the workload allocation scheme as introduced later in Sec.~\ref{sec:hw_space}, all sub-accelerators work on different parts of the feature/weight data in parallel; the hardware utilization and latency can thus be further improved as each sub-accelerator is configured to better fit the corresponding data structure and thus can utilize more parallelism/reuse opportunities.    
\end{itemize}

In particular, the aforementioned template is composed of (1) multiple sub-accelerators which are able to handle both the sparse and dense matrix multiplications, (2) the off-chip memory which holds the data that cannot be stored entirely on chip, and (3) a host CPU to manage the states of different sub-accelerators. Each sub-accelerator has local buffers assigned to the intermediate features (i.e., adjacency matrices), the index for sparse features (assuming a COO format~\cite{Sparsematrix}), and the weights and intermediate outputs, respectively. The buffers among sub-accelerators can be configured to be inter-connected so that the buffered data can be shared to minimize the costly off-chip memory accesses, as presented in Fig.~\ref{fig:hw_template_overview}.

\textbf{The GNN parsing \& sparsity analysis auxiliaries.}
Here we briefly describe GNN parsing \& sparsity analysis auxiliaries for a given GNN, e.g., from PyG~\cite{PyG}.
As shown in Fig.~\ref{fig:overview}, a pre-trained GNN choice is first passed through the blocks of GNN parser and sparsity analyzer before being loaded for G-CoS's automated accelerator generation.
Specifically, (1) the GNN parser helps to extract the GNN structure parameters (e.g., the dimensions of the weights and features) and (2) the sparsity analyzer analyzes the sparsity of each adjacency matrix row. They together help the accelerator design space generator produce all the possible choices and balance the workloads for each sub-accelerator.

\textbf{Flexible workload allocation.} As different GNN structures can have drastically different sparsity patterns and feature/weight dimensions, G-CoS employs \textit{two flexible workload allocation schemes} to ensure each sub-accelerator's assigned workload better fit its micro-architecture, e.g., the processing element (PE) array's dimensions and tiling sizes, to achieve high hardware utilization and thus efficiency. The main allocation principle is that the assigned workload is proportional to each sub-accelerator's capacity which is characterized by its number of PEs.
The two schemes balance the workload with (1) \textit{the number of feature (i.e., adjacency matrix) rows} which will be scaled with the pre-analyzed sparsity of the adjacency matrix and (2) \textit{the number of weight columns}, respectively.

\begin{figure}[t]
    \centering
    \includegraphics[width=0.9\linewidth]{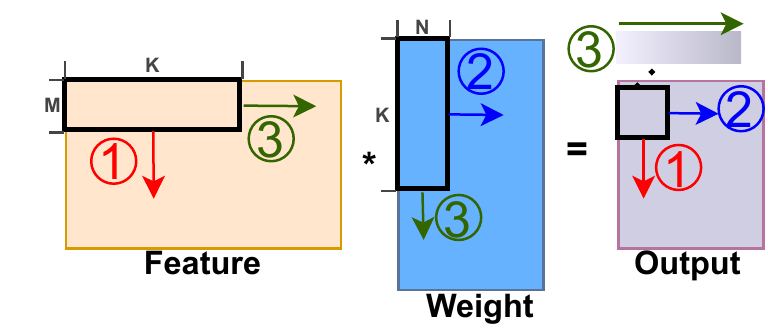}
    \vspace{-1.5em}
\caption{The choices of tiling modes for each sub-accelerator, where the black box represents the temporal tiles, each arrow color represents the first direction that the tile will move towards, and the color gradient represents the output process from start to complete. The three arrow choices favor weight reuses, feature reuses, and output reuses, respectively. For instance, when tiles first move along the red arrow direction, the tiles of the weights can stay stationary. }
    \vspace{-1.5em}
    \label{fig:tiling_order}
\end{figure}

\begin{figure*}[t]
    \centering
    \vspace{-3em}
    \includegraphics[width=\linewidth]{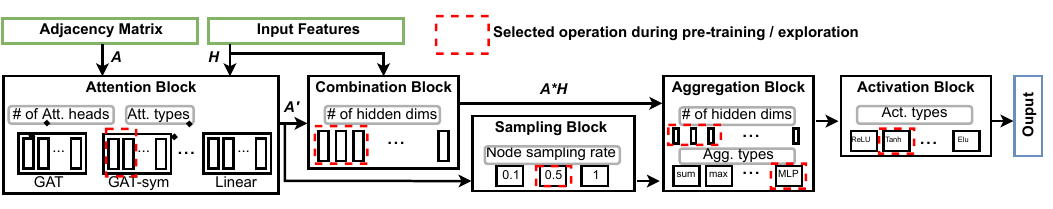}
    \vspace{-2em}
    \caption{The GNN supernet of the proposed G-CoS which covers a comprehensive range of GNN options and is compatible with one-shot NAS methods.}
    \label{fig:supernet}
     \vspace{-1em}
\end{figure*}

\textbf{The sub-accelerator design.} Based on G-CoS's micro-architecture template, the sub-accelerators are auto-generated according to different tiling/kernel modes (as introduced below) to be equipped with different functional components for (1) reflecting different data reuse strategies, (2) favoring different resource trade-offs, and (3) supporting the special operations from various GNN structures, aiming to maximize the hardware efficiency on a wide range of GNNs. Specifically, the sub-accelerators consider:
\begin{itemize}
    \item \textbf{Tiling modes/sizes:} The data per assigned workload may not fit the on-chip memory of each sub-accelerator. As shown in Fig.~\ref{fig:tiling_order}, temporal tiling is then enabled to process the workload temporally. With different tiling modes, a sub-accelerator features more reuses of the features (adjacency matrices), weights, and outputs, respectively. The tiling sizes are defined by K, M, and N (see Fig.~\ref{fig:tiling_order}).
    \item \textbf{Kernel modes:} As illustrated in Fig.~\ref{fig:comp_mode}, with different data mapping and processing patterns for the PE array, each sub-accelerator design would have different off-chip bandwidth consumption, on-chip buffer consumption, and parallelism opportunities.
\end{itemize}

Moreover, each sub-accelerators is equipped with:
\begin{itemize}
    \item \textbf{Dedicated Buffers} to facilitate local reuse opportunities.
    \item \textbf{Dense/SpMM Hybrid Engine} which supports both dense and sparse matrix multiplication within one unit aided with an configurable PE array as in~\cite{shi2021versagnn}.
   \item \textbf{Element-wise Activation Units} to process the non-linear activation operations. 
    \item \textbf{Sampling Units} to schedule the node sampling. 
\end{itemize}
To further increase on-chip reuse opportunities and reduce off-chip accesses, the sub-accelerators can support (1) \textit{weight buffer sharing} which inter-connects all the on-chip weight buffers for weight reuses to reduce off-chip accesses and (2) \textit{buffer re-purposing} where the feature, weight and output buffers are  inter-changeable, so no/reduced off-chip accesses are necessary for the intermediate results between the combination and aggregation phases and/or between layers. Considering the controlling complexity and limited on-chip memory size, if either of these two options are enabled, the number of sub-accelerators will be restricted (e.g., 5 in this work).

\textbf{The searchable accelerator parameters.} Based on the accelerator template (see Fig. \ref{fig:hw_template_overview}), we extract the searchable parameters, of which different combinations lead to different accelerators and form a generic GNN accelerator space to be used by the automated co-search of G-CoS, as summarized in Tab.~\ref{tab:hw_space}. All design choices can be configured differently for each sub-accelerator, except when either \textit{buffer re-purposing} or \textit{weight buffer sharing} is enabled. Otherwise, the tiling and kernel modes are fixed for all sub-accelerators for the ease of controlling and scheduling. The tiling size can range from about 10 to 100 for each sub-accelerator, depending on the given GNN structure. Together, these design choices lead to a hardware design space size of $1$E+$10$ $\sim$ $1$E+$15$.

\begin{table}[!t]
    \vspace{-1em}
    \centering
    \caption{The searchable accelerator parameters for the proposed G-CoS, where $n$ denotes the number of possible tiling sizes ranging about 10$\sim$100 depending on the accelerated GNNs.} 
    \vspace{-0.5em}
    \resizebox{0.5\textwidth}{!}{
    \begin{tabular}{ c||c|c|c|c|c } 
    \hline
    &Tiling & Kernel  & Buffer & WBuf & Tiling  \\
    &Mode   & Mode   & Re-purposing &Sharing & Size\\
    \hline
    \hline
    Choice &\multirow{2}{*}{[0,1,2]} & \multirow{2}{*}{[0,1,2,3]} & \multirow{2}{*}{[0,1]}  & \multirow{2}{*}{[0,1]} & \multirow{2}{*}{[0,...,$n$-1]} \\ 
    Format & &  &  & &  \\
    \hline
    \# of & \multirow{2}{*}{3}  & \multirow{2}{*}{4} & \multirow{2}{*}{2}  & \multirow{2}{*}{2} & \multirow{2}{*}{$n$ ($\sim$10-100)} \\ 
    Choices & &  &  &  & \\
    \hline
    
    \end{tabular}
    }
    \vspace{-2em}
    \label{tab:hw_space}
\end{table}

\subsection{G-CoS: a generic GNAS search space}
\label{sec:gnas_space}

\textbf{The GNN supernet.} To avoid the retraining cost during co-search, G-CoS incorporates a GNN supernet as its GNN design space which is compatible with the adopted one-shot NAS method and able to produce subnetworks covering a comprehensive range of GNN structures, as shown in Fig.~\ref{fig:supernet}. Specifically, the GNN supernet is composed of \textit{five} blocks denoting the attention, combination, sampling, aggregation, and activation blocks. Each block will also have multiple attributes to be determined from a wide range of options as specified in Tab.~\ref{tab:subnet_choices}. For instance, an attention block may assume a GAT-sym structure and have two heads as in Fig.~\ref{fig:supernet}; The combination and aggregation blocks share the same attributes for their hidden dimensions; The attention, sampling, and activation blocks all have a 'skip' option to cover GNNs devoid of these modules. For better generality, each layer of the GNNs assumes this format of supernet, with attribute choices different among layers. For the final/prediction layer, the hidden dimensions and activations are fixed according to the given dataset. Combining all the possible choices in Tab.~\ref{tab:subnet_choices}, the GNN supernet in G-CoS is able to produce $\sim$ $1$E+$9$ choices for a 2 layer GNN, leading to a joint GNN-accelerator space with more than $1$E+$19$ choices.

\textbf{The subnetwork sampling.} The subnetwork is sampled by choosing an option for each attributes of the blocks, e.g., red boxes in Fig.~\ref{fig:supernet}. 
In particular, G-CoS employs uniform random sampling during the pre-training stage, and samples the subnetworks based on the proposed evolutionary algorithm (see Sec.~\ref{sec:G-CoS-alg}) during the exploration stage.

\begin{table}[!t]
    \vspace{-1em}
    \centering
   \caption{The available choices for each block attribute in the GNN supernet, with the attention types (Att. types) following~\cite{gao2019graphnas}. } 
   \vspace{-0.5em}
    \begin{tabular}{ c||c} 
    \hline
   \multirow{2}{*}{ Att. types}& [skip, GCN, GAT, GAT-sym]   \\
      & [COS, Linear, Gene-Linear]\\
    \hline
    Agg. types & [sum, mean, max, MLP] \\
    \hline
   \multirow{2}{*}{ Act. types}& [Skip, Sigmoid, Tanh, ReLu, Linear] \\
              & [Softplus, Leaky ReLu, ReLu6, Elu] \\
    \hline
    \# of hidden &  \multirow{2}{*}{[4, 8, 16, 32, 64, 128, 256]}   \\
    dims    &        \\
    \hline
    \# of Att. &  \multirow{2}{*}{[1,2,4,6,8,16]}   \\
    heads    &        \\
    \hline
        Node  &  \multirow{2}{*}{[0.1,0.5,1]}   \\
    sampling rate    &        \\
    \hline
    \end{tabular}
    \vspace{-2em}
    \label{tab:subnet_choices}
\end{table}

\subsection{G-CoS: one-shot evolutionary GNN-accelerator co-search}
\label{sec:G-CoS-alg}
To tackle the aforementioned challenge of excessively large GNN-accelerator joint search space and costly retraining rooted in many NAS methods~\cite{gao2019graphnas,kyriakides2021evolving}, we propose to employ an one-shot based search approach as inspired by~\cite{guo2020single}, to decouple the supernet pre-training and GNN-accelerator co-search processes, along with an evolutionary algorithm to efficiently navigate through the large joint space to locate optimal GNN-accelerator design pairs for boosting both task accuracy and hardware efficiency. In particular, we only pre-train the GNN supernet once and only inference on the validation set as needed during the exploration stage. To the best of our knowledge, we are the first to study the effectiveness of one-shot NAS within the scope of GNNs. 

\textbf{The supernet pre-training.}
During supernet pre-training, a random subnetwork is uniformly sampled from the GNN supernet by selecting the attribute options from each block and then the subnetwork weights $\omega$ are updated via back-propagation. As~\cite{guo2020single} pointed out, uniform sampling can decouple the weights among possible subnetworks and thus provide a better estimate for their individual fully trained accuracy when these subnetworks are inferenced on the validation dataset during the GNN structure exploration.

\textbf{The weight sharing.}
For more effective pre-training, G-CoS adopts a weight sharing strategy as inspired by~\cite{guo2020single,gao2019graphnas} during pre-training, such that different subnetworks share certain slices of weights. In particular, the weights for combination and aggregation will be shared and retrieved according to the chosen \# of hidden dimensions when a subnetwork is sampled. For the attention block, only the options belonging to the same attention types share their weights and are retrieved according to the number of attention heads. 

\begin{figure}[t!]
\vspace{-0.5em}
    \begin{minipage}{0.47\textwidth}
        \begin{algorithm}[H]
            \caption{G-CoS's GNN-accelerator co-search algorithm}
            \label{alg:ea}
            \begin{algorithmic}[1]
                \footnotesize
                \STATE{ {\bfseries Inputs:}  the target performance $T$;  the number of outputs $N_2$; the samples pool size $p_{max}$, the fitness function $\text{fit}()$; the mutation function $\text{mut}()$; the birth/dying rate $s$}
                
                \STATE{ {\bfseries Outputs:}  $N_2$ number of best found designs $O_{N_2}$}
               \STATE {\bfseries [Procedures]:}
                \STATE{$P$ =\{\}; $fit_{avg}=0$ }
                \STATE{\textbf{While} $fit_{avg} < T$}
                    \STATE{\hspace{8pt} \textbf{If}  $|P|$ $\leq$ $p_{max}$}
                        \STATE{\hspace{16pt} \textbf{If} $|P|$ == $0$}
                            \STATE{\hspace{24pt} Randomly generate $(s*p_{max})$ new designs $pnew_{[gnet, hw]}$}
                        \STATE{\hspace{16pt} \textbf{Else}}
                            \STATE{\hspace{24pt} Find top $(s*p_{max})$ fit designs $top_{[gnet, hw]}$}
                            \STATE {\hspace{24pt} Mutate for new designs $pnew_{[gnet, hw]}$ = mut( $top_{[gnet, hw]}$)}
                    \STATE{\hspace{16pt} Evaluate new designs for fitness:  fit($pnew_{[gnet, hw]}$) }
                    \STATE{\hspace{16pt} Add \{$pnew_{[gnet, hw]}$, fit($pnew_{[gnet, hw]}$)\} to $P$}
                    \STATE{\hspace{8pt} \textbf{Else}}
                        \STATE{\hspace{16pt} Remove the bottom $s*p_{max}$ designs from $P$ }
                    \STATE{\hspace{8pt} $fit_{avg}$ = average fitness of top $N_2$ designs in $P$}
                    \STATE{\textbf{Return} $N_2$ top fit designs  $O_{N_2}$}
            \end{algorithmic}
        \end{algorithm}
    \end{minipage}
\vspace{-2em}
\end{figure}

\textbf{The evolutionary co-search algorithm.} As illustrated in Alg.~\ref{alg:ea}, a specially tailored evolutionary algorithm is developed to efficiently search for best satisfied GNN-accelerator pairs, characterized by $(gnet, hw)$. Overall, it operates by constantly generating new designs around the good design options stored in a pool $P$, filtering out the inferior designs, and then outputting the top $N_2$ performing designs. Specifically, the algorithm inputs include: (1) the fitness function fit(), which evaluates the designs given the GNN-accelerator specs $(gnet, hw)$, (2) mutation function mut() which randomly changes the attributes of good designs to generate new designs, (3) the largest samples pool size $p_{max}$, (4) the birth/dying rate which controls how many designs will be generated/filtered out, and (5) performance target $T$ which determines the terminating conditions and follows the same units as the fitness function. After search, the algorithm outputs the top $N_2$ performing designs within $P$. More input specification choices are provided in Sec.\ref{sec:exp_setup}.


%% file: Sections/4-Experiment.tex
\section{Experiment Results}

In this section, we first introduce the experiment setups in Sec. \ref{sec:exp_setup}, and then benchmark the proposed G-CoS with SOTA GCN accelerators, GNAS and handcrafted GNNs in Sec.~\ref{sec:overall_comp}, Sec.~\ref{sec:against_GNAS} and Sec.~\ref{sec:comp_hand}, respectively. 
\vspace{-0.5em}

\subsection{Experiment Setup}
\label{sec:exp_setup}
\textbf{Baselines and datasets.}
For evaluating G-CoS  \ul{over SOTA GNN accelerators}, we consider three baselines: HyGCN \cite{yan2020hygcn}, AWB-GCN \cite{geng2020awb}, 
and Deepburning-GL \cite{liang2020deepburning}, respectively. For evaluating G-CoS \ul{over SOTA GNAS}, we consider three GNAS baselines: GraphNAS~\cite{gao2019graphnas}, Auto-GNN~\cite{zhou2019auto}, and AutoGraph~\cite{kyriakides2021evolving}. For benchmarking \ul{over SOTA handcrafted GNNs}, we consider four baselines: GCN~\cite{kipf2017semi}, GAT~\cite{GAT}, LGCN~\cite{gao2018large}, and GraphSAGE~\cite{hamilton2017inductive}, \textcolor{black}{covering the most common GNN variants}. Our experiments are conducted on \ul{four datasets}: three citation graph datasets (Cora, CiteSeer and Pumbed) \cite{sen2008collective}, and the Reddit post dataset \cite{hamilton2017inductive}), respectively.

\textbf{GNN training setup.}
For the GNN training, we follow the same dataset splits as  \cite{kipf2017semi,hamilton2017inductive,hu2020ogb}.
The GNN supernet is trained using an Adam optimizer \cite{kingma2014adam} with a learning rate of 0.001 for 1000 epochs, L2 regularization, and dropout. After the design space exploration is finished, the final derived models are trained with additional 400 epochs from scratch under the same configurations. 

\textbf{Evolutionary search algorithm setup.}
For the evolutionary search algorithm presented in Alg.~\ref{alg:ea}, we use a set of generic configurations. Specifically, the fitness function is set as the weighted sum of (inverse)latency and task accuracy, such that latency and task accuracy contributes similarly to the fitness score. The mutation function is set to have a 50\% chance, i.e., the selected GNN-accelerator design pairs have half of the randomly picked design attributes changed. \textcolor{black}{The birth/dying rate ($s$) is set to 0.2.} 

\textbf{Hardware experiment setup.}
To evaluate G-CoS's generated accelerators, we adopt standard the FPGA evaluation and implementation flows in Vivado 2020.2~\cite{vivado}. For the platform, we picked the Xilinx VCU128 FPGA board~\cite{vcu128}, which is equipped with 9024 DSPs, 42MB on-chip memory and 460GB/s HBM of-chip memory. For a fair comparison with other baselines, we limit the DSP consumption to be less than 4096 throughout the design. The generated GNN accelerators are clocked at 330MHz and adopt a 16-bit fixed point precision. For the accelerator template, we fix the number of sub-accelerators to 5 (unless otherwise specified). Note that the design principles of G-CoS introduced in Sec.~\ref{sec:hw_space} is platform agnostic, i.e., G-CoS can be flexibly extended to other platforms such as ASIC. During the design space exploration, to quickly go over numerous design options, we design and implement an \textcolor{black}{in-house performance} simulator to measure the execution time (i.e., the number of cycles).

\begin{figure*}[!t]
    \vspace{-3.5em}
    \centering
    \includegraphics[width=0.9\linewidth]{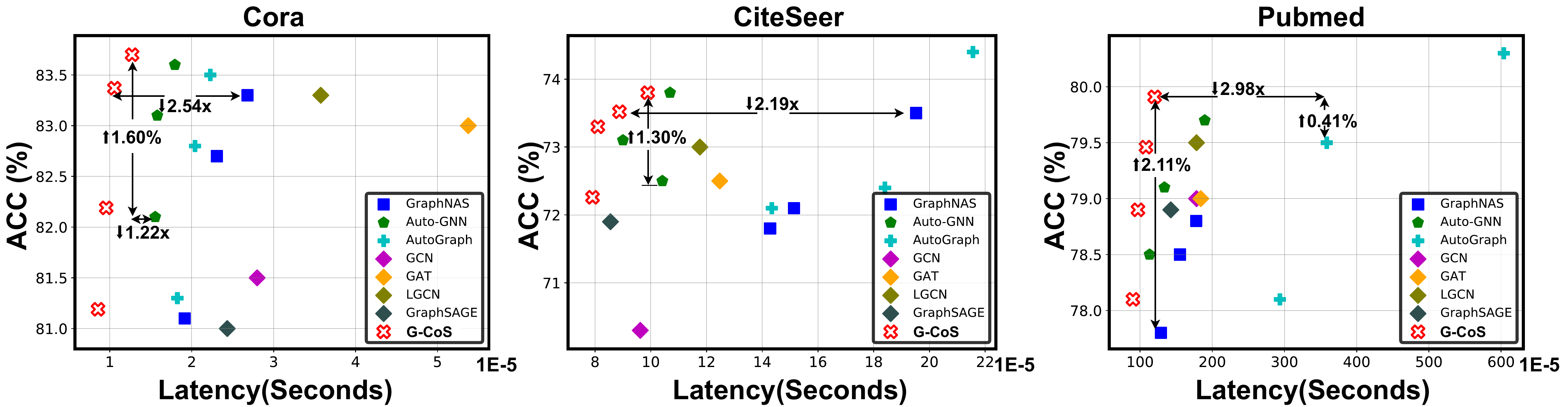}
    \vspace{-1em}
    \caption{Accuracy vs. Latency of G-CoS over the SOTA GNAS and handcrafted GNNs across Cora, CiteSeer and Pubmed datasets. }
    \label{fig:acc_latency}
    \vspace{-1.4em}
\end{figure*}

\begin{figure*}[!t]
    \centering
    \includegraphics[width=1\linewidth]{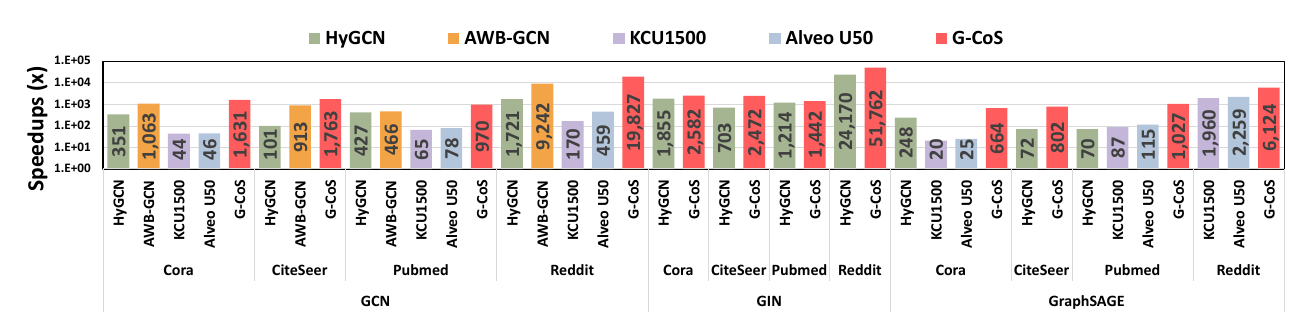}
    \vspace{-3em}
    \caption{The normalized inference speedups (w.r.t. PyG-CPU) achieved by G-CoS over the four SOTA baseline platforms on three GNN models and four representative graph datasets, where KCU1500 and Alveo U50 are two implementation platforms for Deepburning-GL~\cite{liang2020deepburning}}
    \label{fig:hw_perf_1}
    \vspace{-1.8em}
\end{figure*}

\subsection{G-CoS over SOTA GNN accelerators}
\label{sec:overall_comp}
In this set of experiments, we compare G-CoS with existing SOTA GNN accelerators: HyGCN \cite{yan2020hygcn}, AWB-GCN \cite{geng2020awb}, and Deepburning-GL \cite{liang2020deepburning}, in terms of latency and bandwidth consumption. For a fair comparison, we fix the GNN structures and datasets in G-CoS to be the same as the baselines and only search for the accelerator parameters. Since most of them do not provide absolute performance values while reporting the relative speedups over PyG-CPU on an Intel Xeon E5-2680 v3 CPU instead. We also measure and verify the latency on the same CPU \textcolor{black}{and calculate the FPGA speedups over it}, so that PyG-CPU is a common baseline for all methods, and then we can analyze the relative improvements as elaborated below: 

\underline{(1) Speedup}. 
As shown in Fig. \ref{fig:hw_perf_1}, G-CoS achieves an average of 5.52$\times$, 1.92$\times$, 35.98$\times$, and 21.54$\times$ speedups over HyGCN, AWB-GCN, Deepburning-GL-KCU1500, and Deepburning-GL-Alveo U50, respectively. G-CoS's much reduced latency is mostly attributed to its enabling better hardware(DSP) utilization as the multi-sub-accelerator scheme of G-CoS can better cover GNNs' high irregularity and the wide range of searchable accelerator parameters make G-CoS's searched accelerators matching different variation in GNNs. 
\underline{(2) Off-chip memory Bandwidth Consumption}. G-CoS  only requires an average of 50\% off-chip memory bandwidth as compared to HyGCN. The high bandwidth of HyGCN is due to to its high-degree parallelism while G-CoS's versatile kernel mode as illustrated in Sec.~\ref{sec:hw_space} alleviates such an off-chip bandwidth pressure.
\vspace{-1em}
\subsection{G-CoS over SOTA GNAS.}
\label{sec:against_GNAS}
In this set of experiments, we benchmark G-CoS with the SOTA GNAS works: GraphNAS~\cite{gao2019graphnas}, Auto-GNN~\cite{zhou2019auto}, and AutoGraph~\cite{kyriakides2021evolving} based on the metrics of task accuracy and latency. For G-CoS, we co-optimize both the GNN and accelerator design parameters as introduced in Sec.~\ref{sec:hw_space} and Sec.~\ref{sec:gnas_space}, respectively. For a fair comparison, we also accelerate the baselines' generated GNNs under the same hardware platform and optimize the corresponding accelerator parameters. To demonstrate the tradeoff between the hardware efficiency and task accuracy, we restricted the baseline generated GNNs with different flops and then select the designs with the lowest latency; for G-CoS, we simply decrease/increase the weighting coefficient of the latency in the search metrics to achieve flexible tradeoffs.  The results are presented in Fig.~\ref{fig:acc_latency}: G-CoS consistently maintains a better performance frontier, i.e., a higher accuracy and lower latency. In particular, G-CoS achieves a 1.60\%, 1.3\% and 2.11\% increase in task accuracy with a similar or lower latency, and a 2.54$\times$, 2.15$\times$ and 2.98$\times$ latency reduction with a similar or higher accuracy, as compared with the SOTA GNAS works on the Cora, CiteSeer and Pubmed datasets, respectively. Although, considering flops during the search can offer some guidance for more hardware-friendly GNNs, it can not fully capture the compatibility between the searched GNNs and the specific hardware platform, resulting the searched GNNs which satisfy the flops requirement but might be hard to accelerate. Thus, co-optimizing the GNN-accelerator pairs can excel both in terms of accuracy and hardware efficiency against traditional GNAS works by offering better guidance and fully customized architecture to every single searched network. 
\textcolor{black}{The entire process takes as low as 4 GPU hours depending on the dataset.}

\vspace{-1em}
\subsection{G-CoS over SOTA handcrafted GNNs.} 
\label{sec:comp_hand}
We also compare the performance of the proposed G-CoS against the SOTA handcrafted GNNs: GCN~\cite{kipf2017semi}, GAT~\cite{GAT}, LGCN~\cite{gao2018large}, and GraphSAGE~\cite{hamilton2017inductive}. For G-CoS, we co-optimize the GNN-accelerator design pairs. For the baselines GNNs, we optimize their accelerator parameters for fair comparison. As shown in Fig.~\ref{fig:acc_latency}, G-CoS's generated GNN-accelerator design pairs consistently achieve better accuracy and lower latency at the same time. Specifically, the G-CoS generated designs achieve up to 2.7\%,  3.22\% and  1.01\%  increase  in accuracy while having 1.91$\times$,  1.08$\times$ and  1.19$\times$ reduction in latency. 


%% file: Sections/5-Conclusion.tex
\vspace{-1em}
\section{Conclusion }

We propose G-CoS, a GNN and accelerator co-search framework to automatically search for matched GNN structures and accelerators to maximize both task accuracy and acceleration efficiency. To the best of our knowledge, G-CoS is the first co-search framework for GNNs and their accelerators. Extensive experiments show that the GNNs and accelerators generated by G-CoS consistently outperform SOTA GNNs and GNN accelerators, while only requiring a few hours for the end-to-end generation of the matched  GNNs and their accelerators. We believe that our G-CoS has made an important heuristic step towards boosted GNN acceleration efficiency and fast development of efficient GNN solutions.

\section*{Acknowledgment}
\vspace{-0.2em}
This work was supported in part by the \textcolor{black}{National Institutes of Health under Award R01HL144683, National Science
Foundation under CAREER-2048183.}